\documentstyle[multicol,prb,aps,epsf] {revtex}
\begin{document}
\bibliographystyle{prbsty}
\draft
\title{The antiferromagnetic order in an F-AF random alternating
quantum spin chain
: (CH$_3$)$_2$CHNH$_3$Cu(Cl$_x$ Br$_{1-x}$)$_3$
}
\author{Tota Nakamura}
\address{Department of Applied Physics, Tohoku University,
         Aoba, Sendai, 980-8579, Japan }
\date{\today}
\maketitle
\begin{abstract}
A possibility of the uniform antiferromagnetic order is pointed out
in an $S=1/2$ ferromagnetic (F) - antiferromagnetic (AF)
random alternating Heisenberg quantum spin chain compound:
(CH$_3$)$_2$CHNH$_3$Cu(Cl$_x$ Br$_{1-x}$)$_3$.
The system possesses the bond alternation of strong random 
bonds that take $\pm 2J$ and weak uniform AF bonds of $-J$.
In the pure concentration limits, the model reduces to the AF-AF alternation
chain at $x=0$ and to the F-AF alternation chain at $x=1$.
The nonequilibrium relaxation of large-scale quantum
Monte Carlo simulations exhibits critical behaviors of the
uniform AF order in the intermediate concentration region,
which explains the experimental observation of the magnetic phase transition.
The present results suggest that the uniform AF order may survive
even in the presence of the randomly located ferromagnetic bonds.
\end  {abstract}


\begin{multicols}{2}

\narrowtext

Randomness often disorders the orderly, but
sometimes it brings order to the disorderly.
This phenomenon is referred to as {\it order by disorder}.
In the studies of the low-dimensional quantum spin systems
the appearance of magnetic order due to the impurity doping 
of a spin-disordered system has been attracting interest both in 
experimental \cite{hase-azuma-fujiwara}
and theoretical \cite{laukamp-sorensen-tota-Dobry}
research.
The systems commonly possess a good one-dimensionality and 
have a finite energy gap above the nonmagnetic ground state
mostly caused by the singlet dimer state.
The impurity doping may destroy the dimers, creating
magnons, and cause the antiferromagnetic (AF) order.
When the magnetic interactions are all antiferromagnetic,
the appearance of uniform AF order becomes easily recognizable.

Recently, 
Manaka et al.\cite{manaka3} found 
another example of the order by disorder phenomenon.
Two spin-gapped compounds with the nearly identical structure, but
with different origins for their energy gap, were mixed.
One compound is 
(CH$_3$)$_2$CHNH$_3$CuCl$_3$ (IPACuCl$_3$), 
which realizes the F-AF bond
alternation Heisenberg chain
for $J_{\rm strong}\sim$ 54K (F) and $J_{\rm weak}\sim -23$K (AF).
\cite{manaka-FAF}
Because of the strong F bonds the ground state of this compound
is considered the Haldane state.
The other compound mixed is
(CH$_3$)$_2$CHNH$_3$CuBr$_3$ (IPACuBr$_3$), 
which realizes the AF-AF bond alternation Heisenberg chain for
$J_{\rm strong}\sim -61$K (AF) and $J_{\rm weak}\sim -33$K (AF).
\cite{manaka-AFAF}
The ground state is the dimer state.
In the region of intermediate concentration in the mixed compound,
$0.44 < x < 0.87$ in IPACu(Cl$_x$ Br$_{1-x}$)$_3$, 
they observed a magnetic phase transition
through the susceptibility measurement and the specific heat measurement.
Dependences of the susceptibility on the direction of the external 
field suggested that the order is antiferromagnetic, 
but the magnetic structure has not been experimentally clarified yet.

In the mixed compound, 
a kind of the random-bond quantum spin chain is realized.
The randomness is considered to destroy the energy gap, causing 
a magnetic order to appear in the gapless region with the assistance 
of the very weak but finite interchain interactions.
From knowledge of the quantum random spin chain, 
\cite{ma79,fisher94,hyman-hida-hikihara,hyman-y97,hida99}
one finds the uniform order to be totally destroyed when the
F bonds are randomly located.
However, in actuality these F bonds and the AF interchain 
interactions cause randomly-located frustration.
Despite these adverse circumstances the uniform AF order may become critical
in the one-dimensional model.
Particularly, the uniform AF magnetization at the fully random
point decays algebraically
with an exponent that is almost same as the gapless 
$S=1/2$ uniform antiferromagnetic 
Heisenberg (AFH) chain.
This evidence suggests that the uniform AF order
appears in a significant way when the interchain interactions exist.
The gapless phase bounds have also been estimated, and they agree
with the experimental observation of the magnetic phase.
Singular randomness on the strong bonds, $\pm 2J$, and the alternate uniform
bonds are clarified to cause a power-law effective bond distribution.
This distribution is the origin of the critical behavior of the AF order.

Consider the following $S=1/2$ random alternating quantum Heisenberg chain,
\begin{equation}
{\cal H}=-\sum_i ^N
J_{2i-1}
\mbox{\boldmath $S$}_{2i-1}\cdot
\mbox{\boldmath $S$}_{2i  }
+
J_{2i}
\mbox{\boldmath $S$}_{2i  }\cdot
\mbox{\boldmath $S$}_{2i+1},
\end  {equation}
where $J_{2i}=\pm 2J$ indicating the strong random bonds and
$J_{2i-1}=-J$ indicating the weak uniform bonds.
From the crystal structure analyses of the pure compounds
\cite{manaka-FAF,manaka-AFAF}
it is known that
there are two Cl ions or two Br ions that bridge between the Cu ions:
Cu$<^{\rm Cl}_{\rm Cl}>$Cu or
Cu$<^{\rm Br}_{\rm Br}>$Cu.
They are linked stepwisely along the {\it c}-axis.
The exchange interactions within a step are strong, whereas those between
steps are weak.
These has been explained by the overlap of the orbitals of Cu and (Cl, Br).
\cite{manaka-FAF,manaka-AFAF}
The bridging angle may differ between  Cl and Br because of a different
ion radius, causing changes in the exchange interactions.
In the mixed compound,
the exchange interactions on the weak bonds may be insensible to the 
bridging angles because the overlap of the orbitals is small.
Thus, the difference in amplitude on the weak bonds is neglected
(-23K for IPACuCl$_3$ and -33K for IPACuBr$_3$), and instead
they are set to a uniform AF value: $J_{2i-1}=-J$.
On the other hand however,
the sign of the exchange interaction may change
by subtle differences in the bridging angles along the strong bonds
due to the large overlap of the orbitals.
Therefore, the magnitude is set as twice the weak bonds
and the sign is random: $J_{2i}=\pm 2J$.
There are three types of bridging ion configurations:
Cu$<^{\rm Cl}_{\rm Cl}>$Cu
with probability $x^2$, 
Cu$<^{\rm Br}_{\rm Br}>$Cu
with $(1-x)^2$, and 
Cu$<^{\rm Cl}_{\rm Br}>$Cu
with $2x(1-x)$.
It is considered that the Cu$<^{\rm Cl}_{\rm Br}>$Cu configuration yields 
the AF interactions by the following reasoning 
suggested by Manaka.\cite{manakajiba}
Let an F bond concentration (not the Cl ion concentration, $x$)
build up in the random part denoted by $p$.
If one assumes that the
Cu$<^{\rm Cl}_{\rm Br}>$Cu
configuration yields the F interaction,
then $p=x(2-x)$.
The magnetic phase appears for $0.44 \le x \le 0.87$ in the experiment.
This directly corresponds to the
appearance of the dimer phase during $0 \le p \le 0.69$,
and the appearance of the Haldane phase only during $0.98 \le p \le 1$.
This assumption contradicts with our sense that
the Haldane phase is rather robust against randomness compared to
the dimer phase.
Therefore, the 
Cu$<^{\rm Cl}_{\rm Br}>$Cu configuration
and the 
Cu$<^{\rm Br}_{\rm Br}>$Cu
configurations correspond to the AF interactions, and only the 
Cu$<^{\rm Cl}_{\rm Cl}>$Cu 
configuration corresponds to the F interaction.
Then, we get $p=x^2$.

Hyman and Yang \cite{hyman-y97} 
investigated a very similar model wherein the alternate even bonds
are randomly AF and the other bonds are randomly F or AF.
In their model the random singlet phase appears where the
spin correlation decays algebraically as $r^{-2}$.
However, numerical results presented subsequently show that 
the AF magnetization behaves qualitatively the same as the gapless $S=1/2$ 
uniform AFH chain at the fully random point.
Thus, a subtle difference in the bond distributions may produce different
results.
As for the connection with the real materials, 
the random bond distribution of our model
can be realized by the intercalation in the spin-gapped compounds.
Thus, 
the model may account for the various impurity-induced 
magnetic-order phenomena.

In our research
the quantum Monte Carlo with the nonequilibrium relaxation 
analysis \cite{ner-nonomura-shirahata} was used.
Since the nonequilibrium relaxation method essentially handles 
the size-independent relaxation process,
the obtained results can be considered as having an
infinite Trotter number and infinite system size.
In actuality, the nonequilibrium relaxation functions of different values of
the ratio 
$\beta/m$ ($\beta$:the inverse temperature, $m$: the Trotter number)
ride on the same curve when we rescale the Monte Carlo time step
by the correlation time $\tau(m)$ dependent on the Trotter number.
The $m$-dependence only appears due to the finite-size effect.
The value of $\beta/m$ is fixed to $1/2$ in these results.
The simulations start with spin configurations of 
either a quantum ground state of the pure system 
(the dimer state or the Haldane state)
or the classical uniform AF state.
In the case of the quantum state start-up,
the uniform AF susceptibility is observed
whether it diverges algebraically or
converges to a finite value.
When it shows the converging behavior the system is considered to be
in a state of disorder or in the gapful state.
Otherwise, when it shows an algebraic divergence, the AF order is critical 
and the system is in the gapless phase.
The uniform AF magnetization 
is observed whether it decays exponentially or algebraically
when starting from the classical AF state.
The typical size of the system is $N=161$(322 spins), $m=1000$ at 
$\beta J=500$  and the numbers of the random bond configurations are
more than 5000 in the simulations for the quantum states,
and $N=1601$, $m=400$, $\beta J=200$ with several hundreds configurations 
in the simulations for the classical state. 
These sizes of the simulations determine the resolution limit of the
obtained physical quantities.
The resolution of the susceptibility is on the order of $1/T$, which
corresponds to the AF magnetization per spin 
$\langle M_{\rm AF} \rangle \sim 1/\sqrt{mN}\sim 3\times 10^{-3}$.
The resolution of the AF magnetization when starting from the
classical state is on the order of 
$1/\sqrt{mN\times {\rm (number~of~samples)}} \le 10^{-4}$.
Both simulations are consistent when within the resolution of 
$\langle M_{\rm AF} \rangle > 10^{-3}$.
It is difficult to estimate the resolution of the experiment
observing the magnetic phase transition, but if one assumes
$\langle M_{\rm AF} \rangle \sim 10^{-3}$ then
the numerical results presented in this paper
quantitatively explains the experiment.

\begin{figure}[htb]
\begin{center}
 \epsfxsize = 7.0cm
 \epsffile{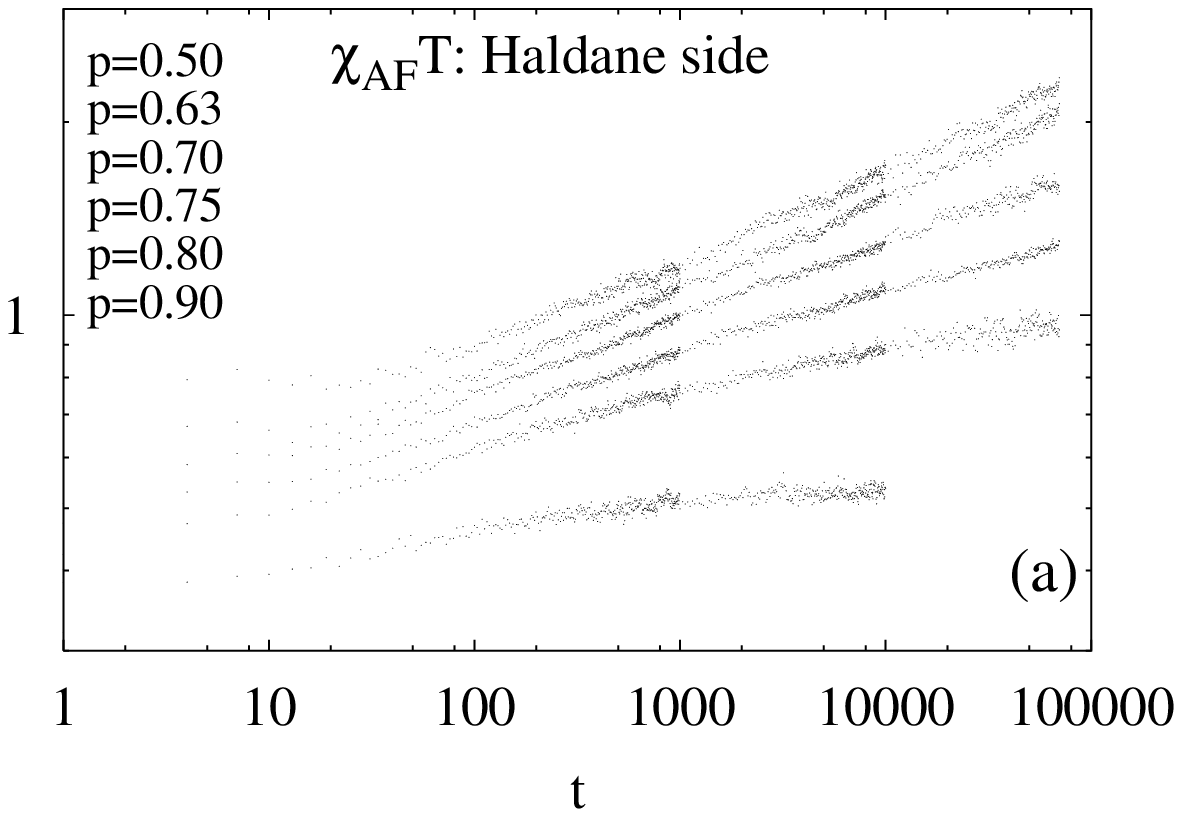}
 \epsfxsize = 7.0cm
 \epsffile{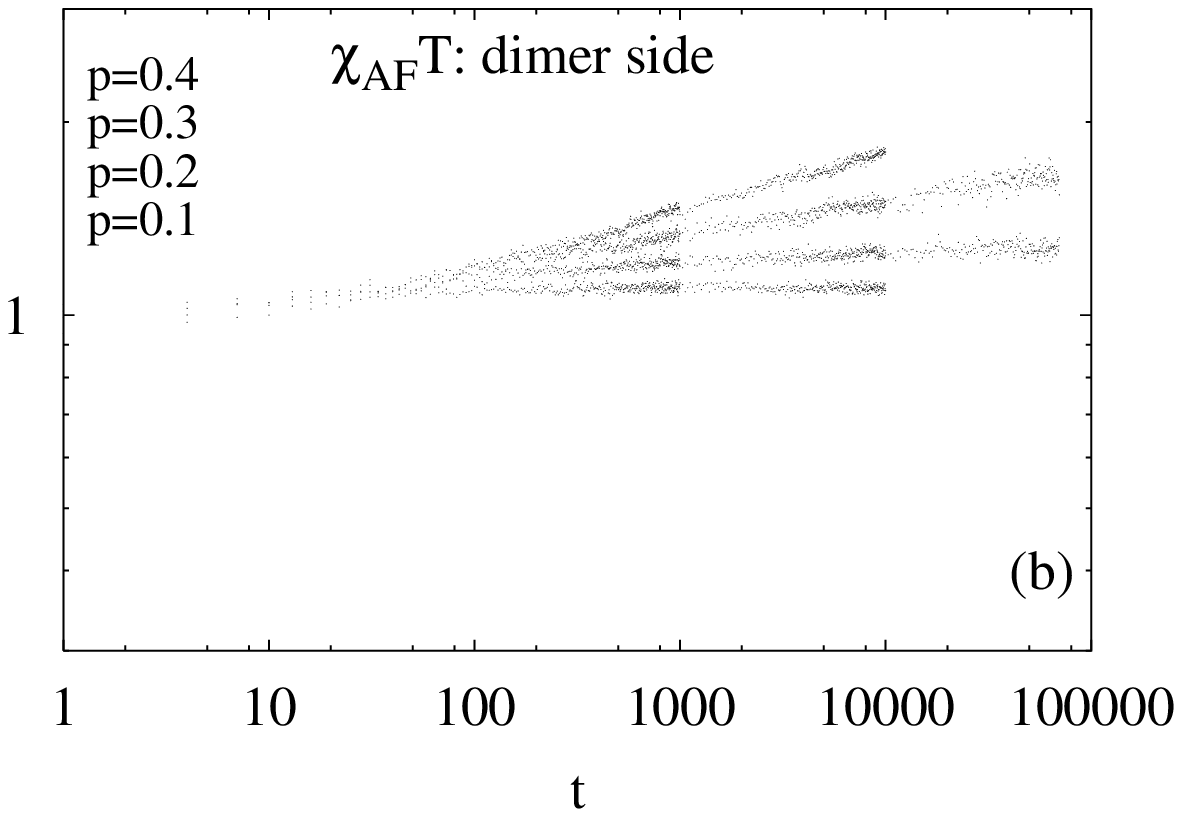}
\end {center}
 \caption {
 Nonequilibrium relaxations of $\chi_{\rm AF}T$.
 The F bond concentrations $p$ values, from top to bottom,
correspond to the data from top to bottom.
 The behavior changes from
 converging to diverging at $p\sim 0.75$ ($x\sim 0.87)$ in the Haldane side and 
 at $p\sim  0.2$ ($x\sim 0.44)$ in the dimer side.
 \label{fig:ust}
          }
\end  {figure}

Figure \ref{fig:ust} shows the nonequilibrium relaxation functions of the
uniform AF susceptibility multiplied by the temperature
in the simulations of the Haldane state (a) and
of the dimer state (b).
After the initial relaxation of 50 steps, the 
susceptibility begins to diverge algebraically in the intermediate region
of $0.2 < p < 0.75$, which corresponds to $0.44 < x < 0.87$ by $p=x^2$.
This displays the critical behavior of the uniform AF order
in the one-dimensional level,
and that the AF order may appear if there exist finite interchain interactions.
The phase boundary coincides quantitatively with
the experimental observation of the magnetic phase, 
and thus the order is speculated to be the uniform AF order.
Of course 
there may be a possibility of other magnetic structures we do not mention here,
such as the $\uparrow \uparrow \downarrow \downarrow$ state in the
Haldane side or the generalized staggered state.
Investigations on these ordered states are left for future study.

\begin{figure}[bht]
\begin{center}
 \epsfxsize = 8.0cm
 \epsffile{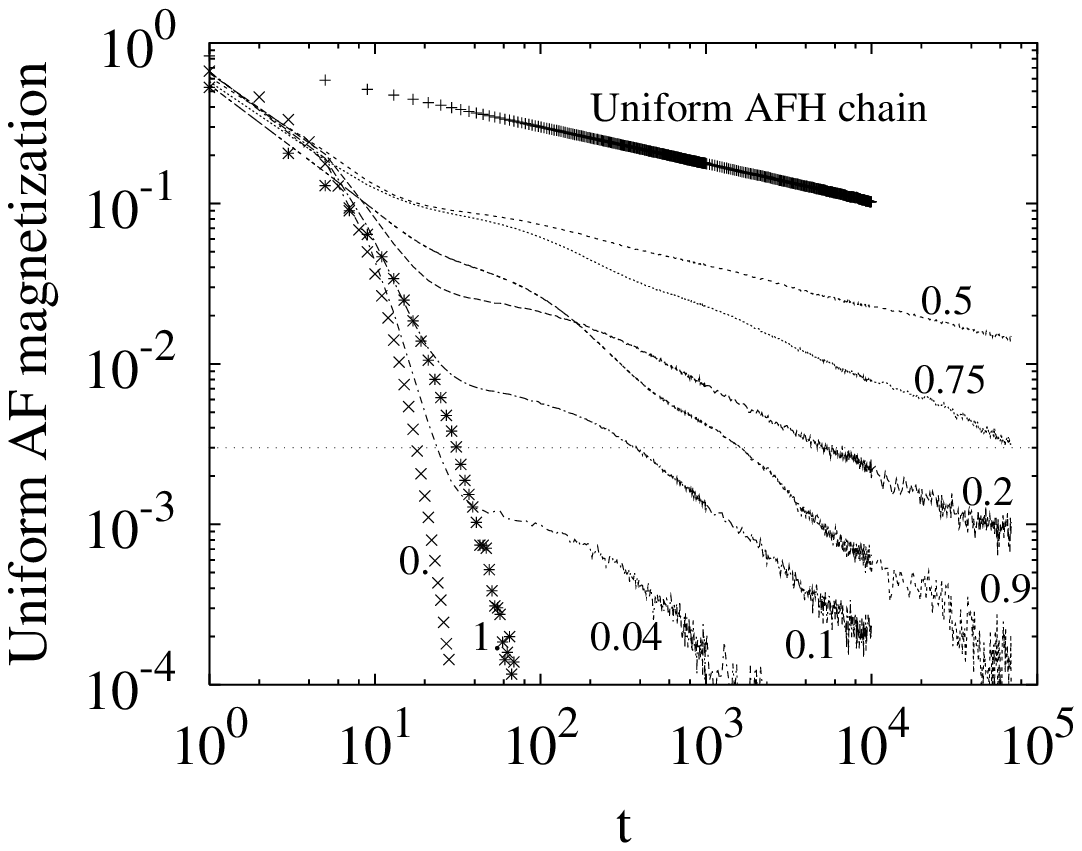}
\end  {center}
 \caption {
Nonequilibrium relaxation of the uniform AF magnetization.
The F bond concentration $p$ is denoted aside each data plot.
At $p=0.5$, the algebraic decay is qualitatively the same 
as the $S=1/2$ uniform AFH chain.
The dotted horizontal line depicts the resolution of
the susceptibility simulation.
 \label{fig:figaf}
}
\end  {figure}

Starting the simulation from the classical AF-ordered state and 
observing the uniform AF magnetization can check the critical nature
of the uniform AF state.
Figure \ref{fig:figaf} shows the nonequilibrium relaxation.
For comparison, the relaxation function
of the pure base system at $p=0$(dimer chain) and
$p=1$ (Haldane chain) and the gapless $S=1/2$ 
uniform AFH chain have been plotted.
In the intermediate region the relaxation function exhibits the critical 
decay after an initial relaxation that rides on the 
relaxation of the pure gapful system.
Note that the length of the initial relaxation, $t<50$, 
agrees with that of the susceptibility simulations in Fig. \ref{fig:ust}.
The numerical value of the AF magnetization at which the relaxation begins
critical behavior is roughly considered as the magnitude of the order.
At $p=0.1$, an algebraic decay begins below 
the susceptibility resolution limit, 
$3 \times 10^{-3}$, and therefore critical behavior was not detected by
the susceptibility simulation nor by the experiment.
Within the present accuracy of the AF-magnetization simulation
the phase boundary in the dimer side resides between $p=0.04$ and $p=0.1$: 
at $x=0.2 \sim 0.3$, which is lower than the experimental observation
$x=0.44$.
When the resolution of the experimental probe is much sharper,
the magnetic phase can be observed in wider region.
For $p > 0.75$, the relaxation shows a multi-exponential type decay,
suggesting a discrete distribution of the energy gap.
However, as the randomness increases the relaxation looks like 
an algebraic decay.
At the fully random point $p=0.5$
the magnetization takes a maximum value.
The slope of the relaxation at this point 
is almost the same as that of the $S=1/2$ uniform AFH spin chain,
but the amplitude is reduced to 1/4.
This behavior guarantees the observation of a weak antiferromagnetic
phase transition assisted by the interchain interactions.

The algebraic decay in the intermediate region
can be explained by the power-law distribution of the energy gap.
In the Haldane side ($p>0.5$), we may consider the strong AF bonds as the
impurity that may be replaced by effective weak bonds (just like the
real-space renormalization procedure).\cite{ma79}
As long as the successive perturbation is good,
the strength of the effective bond replacing $n$ aligned strong AF bonds
are $\exp[-\lambda n]J$ with $\lambda=\log 4$.
The probability of the realization is $(1-p)^n$.
Given these, the distribution of the effective bonds is
\cite{hida01}
\begin{equation}
 P(J)\sim \frac{1}{\lambda}J^{-1+\frac{1}{\lambda}\log\frac{1}{1-p}}
\propto
\alpha J^{-1+\alpha},
\end  {equation}
with $\alpha=(1/\lambda)\log (1/(1-p))$.
Because the VBS picture is valid in this region, the singlet dimers are 
located on the weak AF bonds between the strong F bonds.
This bond distribution is directly interpreted by the gap distribution
$P(\Delta)$, 
and it is equivalent to the distribution in the random singlet phase,
$P(\Delta)=\alpha\Omega ^{-\alpha}\Delta ^{\alpha-1}$, with a characteristic
energy scale $\Omega$.\cite{fisher94}
Since each energy gap contributes to the exponential decay of the
AF magnetization with a contribution $\exp[-\Delta^z t]$
($z$: the dynamic exponent),
the collection of these exponential decays by the distribution $P(\Delta)$
becomes the power-law decay as
\begin{equation}
\int \exp[-\Delta^z t]P(\Delta)d\Delta=\Gamma(\alpha/z+1)(t/\tau)^{-\alpha/z},
\label{eq:gamma}
\end  {equation}
with the characteristic time scale $\tau \sim \Omega^{-z}$.
Therefore, the AF magnetization is expected to decay algebraically with 
the exponent $\alpha/z=(1/z\alpha)\log(1/(1-p))$,
which is dependent on the concentration $p$.
The slope of the algebraic decay in Fig. \ref{fig:figaf}
quantitatively agrees with this expression supposing 
$z=2.2$ for $p>0.6$, and $z=2.0$ for $0.5 \le p< 0.6$.
In the dimer side ($p<0.5$), the interpretation is not straightforward.
It is not good to deplete the strong AF bonds which are the majority.
Depletion of the strong F bonds (the Haldane cluster) is possible, and
weak effective bonds of the amplitude $\exp[-\lambda' n]$ \cite{hagiwara90}
may replace them.
This way we can obtain the same critical behavior.
However, the estimated values of the exponent do not 
coincide with the numerical results 
of Fig. \ref{fig:figaf}
regardless of its success on the Haldane side.
For the quantitative agreements, we need further interpretation.

\begin{figure}[bht]
%\begin{center}
 \epsfxsize = 8.0cm
 \epsffile{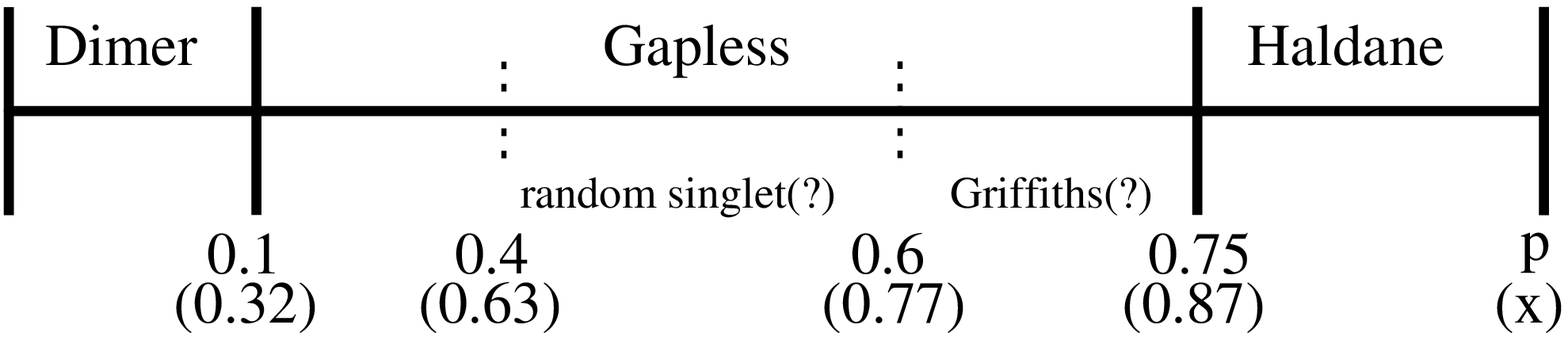}
%\end  {center}
 \caption {A rough speculation of a phase diagram.
 \label{fig:phase}
}
\end  {figure}

Theoretical speculations on the phase boundary are possible.
It is pointed out by Hida \cite{hida99} that the Haldane phase is 
stable as long as the gap distribution is not singular at $\Delta=0$.
In the present model, it corresponds to $p=0.75$ by $\alpha=1$.
This value is consistent with the numerical results of both 
Fig. \ref{fig:ust} and Fig. \ref{fig:figaf}.
It is also commented 
by Yang and Hyman \cite{yang-h2000}
that the random singlet phase begins at $\alpha\sim 0.67$
for the algebraic bond distribution $P(J)\sim \alpha J^{-1+\alpha}$.
This point corresponds to $p\sim 0.6$, and in actuality
the dynamic exponent $z$ seems to change the value at this point from
$z=2.2$ to $z=2$.
In the neighborhood of the fully random point, $0.4 < p < 0.6$,
the exponent is weakly dependent on $p$ suggesting a universal phase.
However, the critical behavior, like the uniform AFH chain,
suggests that the spin correlation decays
as $r^{-1}$ not as $r^{-2}$ in the random singlet phase.
This point is still unclear, and further investigations are necessary.
In our preliminary simulations 
the relaxation of the string order parameter 
suggests that the string order is more than critical (ordered or critical) 
for $0.6 < p \le 1$. 
Recently, Damle \cite{damle} argued the existence of the Griffiths phase
in the random AFH S=1 chain,
where the density of states have power-law distributions and the 
dynamical exponent depends on the model parameter.
The similarities of the model and the obtained physical results suggest
that the Griffiths phase might exist in the region of $0.6 < p < 0.75$. 
Regardless,
the uniform AF order becomes critical in the gapless region,
$0.1 < p < 0.75$, as is observed in the experiment.
A speculated phase diagram is shown in Fig. \ref{fig:phase}. 
This phenomenon is quite interesting because the uniform AF order
survives despite the randomly located ferromagnetic bonds.
This may mean the role of the ferromagnetic bonds in the 
quantum system should be reconsidered.

In this paper
the magnetic ordered phase observed in 
the experiment of IPACu(Cl$_x$~Br$_{1-x}$)$_3$ is explained
and the magnetic structure is speculated to be the uniform AF order.
The phase boundaries estimated in the simulation quantitatively agree with
the experimental results.
Behaviors of the relaxation function in Fig.~\ref{fig:figaf} suggest
the following scenario for the appearance of the AF order.
Short-range AF correlations are first destroyed by local dimer states
or local Haldane states that appear during the initial relaxation of
the exponential decay.
Beyond these local states there exist very weak but finite
effective interactions that have the power-law distribution.
Collections of the exponential decay due to these effective interactions
becomes the algebraic decay seen in Eq. (\ref{eq:gamma}).
This is considered as the origin of the criticality of the uniform
AF order, and with help of the interchain interaction the magnetic
phase is observed in the experiment.
The present mechanism may explain, and thus serve as a new interpretation for,
general impurity-induced long-range order phenomena.

The author would like to thank Dr. H. Manaka and Professor 
K. Hida for fruitful 
discussions and comments.
He also thank A. Baldwin for reading the manuscript.
Use of the random number generator RNDTIK programmed by
Professor N. Ito and Professor Y. Kanada is gratefully acknowledged.

\end{multicols}


\begin{thebibliography}{99}

\bibitem{hase-azuma-fujiwara}
  M. Hase et al.,
  Phys. Rev. Lett. {\bf 71}, 4059 (1993);
  M. Azuma et al.,
  Phys. Rev. B {\bf 55}, R8658 (1997);
  N. Fujiwara et al.,
  Phys. Rev. Lett. {\bf 80}, 604 (1998).

\bibitem{laukamp-sorensen-tota-Dobry}
  M. Laukamp et al.,
  Phys. Rev. B {\bf 57}, 10755 (1998);
  E. S\o rensen et al.,
  Phys. Rev. B {\bf 58}, R14701 (1998);
  T. Nakamura, Phys. Rev. B {\bf 59}, R6589 (1999);
  A. Dobry et al.,
  Phys. Rev. B {\bf 60}, 4065 (1999).

\bibitem{manaka3}
  H. Manaka, I. Yamada, and H. Aruga Katori,
  Phys. Rev. B {\bf 63}, 104408 (2001).

\bibitem{manaka-FAF}
H. Manaka, I. Yamada, and K. Yamaguchi,
J. Phys. Soc. Jpn. {\bf 66}, 564 (1997);
H. Manaka, I. Yamada, Z. Honda, H. Aruga Katori, and K. Katsumata,
{\it ibid.} {\bf 67}, 3913 (1998).

\bibitem{manaka-AFAF}
H. Manaka and I. Yamada,
J. Phys. Soc. Jpn. {\bf 66}, 1908 (1997).

\bibitem{ma79}
  S. K. Ma, C. Dasgupta, and C-K. Hu, Phys. Rev. Lett. {\bf 43}, 1434 (1979).

\bibitem{fisher94}
  D. S. Fisher, Phys. Rev. B {\bf 50}, 3799 (1994).

\bibitem{hyman-hida-hikihara}
  R. A. Hyman, K. Yang, R. N. Bhatt, and S. M. Girvin,
  Phys. Rev. Lett. {\bf 76}, 839 (1996);
  K. Hida,
  J. Phys. Soc. Jpn. {\bf 66}, 330 (1997);
  T. Hikihara, A. Furusaki, and M. Sigrist,
  Phys. Rev. B {\bf 60}, 12116 (1999).

\bibitem{hyman-y97}
  R. A. Hyman and K. Yang,
  Phys. Rev. Lett. {\bf 78}, 1783 (1997).

\bibitem{hida99}
  K. Hida,
  Phys. Rev. Lett. {\bf 83}, 3297 (1999).

\bibitem{manakajiba}
 H. Manaka, I. Yamada, H. Mitamura, and T. Goto, unpublished.

\bibitem{ner-nonomura-shirahata}
N. Ito, Physica A {\bf 196}, 591 (1993);
N. Ito, and Y. Ozeki, Intern. J. Mod. Phys. {\bf 10}, 1495 (1999);
Y. Nonomura, J. Phys. Soc. Jpn.  {\bf 67}, 5 (1998);
T. Shirahata and T. Nakamura, Phys. Rev. B {\bf 65}, 024402 (2002).

\bibitem{hida01}
K. Hida, cond-mat/0111521.

\bibitem{hagiwara90}
M. Hagiwara et al.,
Phys. Rev. Lett. {\bf 65}, 3181 (1990).

\bibitem{yang-h2000}
K. Yang and R. A. Hyman, Phys. Rev. Lett. {\bf 84}, 2044 (2000).

\bibitem{damle}
K. Damle, cond-mat/0201118.
\end  {thebibliography}
\end{document}